\documentclass[useAMS,usenatbib]{mn2e}
\usepackage{graphicx}

\newcommand{\pasa}{PASA}

\newcommand{\mnras}{MNRAS}

\def\lsim{\raise0.3ex\hbox{$<$}\kern-0.75em{\lower0.65ex\hbox{$\sim$}}} 
\def\gsim{\raise0.3ex\hbox{$>$}\kern-0.75em{\lower0.65ex\hbox{$\sim$}}} 

\title[Primordial magnetic field constraints from the end of reionization]{Primordial magnetic field constraints from the end of reionization}
\author[Dominik R. G. Schleicher and Francesco Miniati]{Dominik R. G. Schleicher$^{1}$\thanks{E-mail:
dschleic@astro.physik.uni-goettingen.de} and Francesco Miniati$^{2}$\thanks{E-mail: fm@phys.ethz.ch}\\
$^{1}$Georg-August-Universit\"at, Institut f\"ur Astrophysik, Friedrich-Hund-Platz 1, 37077 G\"ottingen, Germany\\
$^{2}$Physics Department, Wolfgang-Pauli-Strasse 27, ETH-Z\"urich, CH-8093 Z\"urich, Switzerland}
\begin{document}


\pagerange{\pageref{firstpage}--\pageref{lastpage}} \pubyear{2002}

\maketitle

\label{firstpage}

\begin{abstract}
  Primordial magnetic fields generated in the early universe are
  subject of considerable investigation, and observational limits on
  their strength are required to constrain the theory.  Due to their
  impact on the reionization process, the strength of primordial
  fields can be limited using the latest data on reionization and the
  observed UV-luminosity function of high-redshift galaxies.  Given
  the steep faint-end slope of the luminosity function, faint galaxies
  contribute substantial ionizing photons, and the low-luminosity
  cutoff has an impact on the total budget thereof.  Magnetic pressure
  from primordial fields affects such cutoff by preventing collapse in
  halos with mass below $10^{10}$~M$_\odot (B_0 / 3~nG)^3$, with $B_0$
  the co-moving field strength. In this letter, the implications of
  these effects are consistently incorporated in a simplified model
  for reionization, and the uncertainties due to the cosmological
  parameters, the reionization parameters and the observed UV
  luminosity function are addressed. We show that the observed
  ionization degree at $z\sim7$ leads to the strongest upper limit of
  $B_0\lsim 2-3$~nG. Stronger limits could follow from measurements of high
  ionization degree at $z>7$.
\end{abstract}


\section{Introduction}

The origin of magnetic fields in the universe is subject of active
study. Magnetic fields are a common feature of nearby and distant 
galaxies~\citep{Beck96,Bernet08,Murphy09}, where they quickly build 
up due to small scale dynamo action~\citep[e.g.][]{Schleicher10c,Sur10, FederrathPRL, Schober11}.
Magnetic fields exist also
in the intergalactic medium (IGM). 
In particular, galaxy clusters exhibit $\mu$G
fields~\citep{Clarke01} which require non-negligible initial
seeds~\citep{Dubois08,Ryu08,Miniati11b}.
Also, tiny but significant magnetic fields appear to exist in
cosmic voids, as recently suggested by gamma-ray
experiments~\citep[][but see \citet{2011arXiv1106.5494B}]{Neronov10,tavecchio10,Taylor11}. The generation of magnetic fields in cosmic environments is
generally slower than in galaxies, owing to the longer dynamical
timescales.

Various astrophysical mechanisms have been proposed for the origin of
intergalactic magnetic
fields~\citep{Miniati11,Schlickeiser03,bertone06,Ando10}.  In
particular,~\citet{Miniati11} make consistent
predictions for the magnetic fields observed in the cosmic voids.
Primordial models of magnetogenesis~\citep[e.g.][]{Grasso01} provide
an alternative scenario. Unlike most astrophysical cases, their seeds
are not-necessarily negligible.  This is particularly so {\bf for}
some inflationary scenarios \citep{Turner88}, electroweak phase
transition \citep{Baym96}, or QCD phase transition \citep{Quashnock89,
  Cheng94, Sigl97, Banerjee03}. Therefore several efforts to constrain
primordial seeds have been made.

Probing magnetic fields at high redshift is however still
challenging. A primordial magnetic field has an impact on the CMB
anisotropies~\citep[e.g.][]{Barrow97,Subramanian06, Durrer07} and the
latest analysis implies an upper limit on the co-moving field strength
$B_0 < $3nG at 95\% CL~\citep{Yamazaki10}, where $B_0\equiv
B(z)/(1+z)^2$.  Constraints from Big Bang nucleosynthesis can also be
obtained, with $B_0\lsim 1\mu$G~\citep{Grasso96}.  {Considerations
  of gravitational wave production seem to rule out magnetic field
  production during primordial phase transitions \citep{Caprini09}.}
Finally, as pointed out { in various works \citep[e.g.][]{Coles92, Kim96, Battaner97, Sethi05, Tashiro06a}},
the magnetic field can affect the evolution of the IGM
and the growth of structure. These effects can be used to set independent
constraints on primordial fields during
reionization~\citep{SchleicherBanerjee08}.  In particular, the
magnetic field adds a pressure term which affects the
Jeans mass. Also, if the magnetic field is not parallel to the
electric current, charges are subject to a Lorentz {\bf force}.
This gives rise to ambipolar diffusion of the charges
through the neutrals and heating of the IGM
due to collisions between the two species.

In this letter, we constrain primordial magnetic fields by considering
their impact on the reionizaton process. Our analysis
is based on the latest data of the high redshift IGM.
In particular, we use observed UV luminosity function at
redshifts $z\sim4-8$ \citep{Bouwens10, Bouwens11}
and the recent observations of a $z=7.085$
quasar \citep{Mortlock11} which indicates a fraction of neutral hydrogen
of $10^{-4}-10^{-3}$ at $z\sim7$ \citep{Bolton11}.
We find that $B_0\lsim 2-3$nG, similar to
findings from the latest CMB  analysis~\citep{Yamazaki10}.
More importantly, our constraint is complementary with the latter as it is
based on different physical processes and refers to a different
cosmological epoch. Throughout we use the cosmological parameters
from WMAP7~\citep{Komatsu11}.

\section{Technical approach}
In order to assess the impact of primordial magnetic fields on
reionization we solve the following equation 
for the evolution of the volume fraction of ionized hydrogen, 
$Q_{HII}$~\citep{Madau99}
\begin{equation} \label{dqdt:eq}
\frac{dQ_{HII}}{dt}=-\frac{Q_{HII}}{t_{rec}}+
\frac{{\rm SFR}(z)f_{esc}10^{53.2}}{n_H(0)},\label{reion}
\end{equation}
where $n_H(0)$ is the comoving number density of neutral hydrogen
and the other parameters are described below. 
For our reference model we assume the parameter values
as in~\citet[][see also references therein]{Bouwens11},
which reproduce the Thomson optical depth observed by WMAP7 within
$1\sigma$. However, for the cosmological parameters we use the more recent
WMAP7 results~\citep{Komatsu11}. As discussed below, this makes 
a negligible difference. So,
the adopted escape fraction of Lyman-continuum photons is
$f_{\mathrm{esc}}=0.2$. The temperature of the 
reionized IGM is  $2\times10^4$~K, leading to
an hydrogen recombination timescale
\begin{equation}
t_{\mathrm{rec}}=1.0~\mathrm{Gyr} \left( \frac{1+z}{7}\right)^{-3}C_3^{-1},
\end{equation}
where $C_3$ is the clumping factor normalized to a value of $3$. 
Note that the recombination timescale
would decrease by a factor of $2$ for an IGM temperature of
$10^4$~K. The factor  $10^{53.2}$ in Eq.~(\ref{dqdt:eq})
is the production rate of Lyman-continuum photons per $M_\odot$yr$^{-1}$
of forming stars, enhanced by $30\%$ for metal-poor stars in galaxies
in the early universe~\citep{Bouwens11}.
The cosmic star formation rate, SFR,
uncorrected for dust attenuation and in $M_\odot$yr$^{-1}$, is
\begin{equation} \label{SFR:eq}
{\rm SFR}=\int_{-\infty}^{M_{UV}^\mathrm{max}} \Phi(M_{UV})  \dot\rho_*(M_{UV})  d\left( 10^{-0.4(M_{UV}-M_{UV}^*)} \right),\nonumber
\end{equation}
where $\Phi(M_{UV})$ is the Schechter function as a function of
the UV magnitude, $M_{UV}$, and $M_{UV}^*$ its characteristic magnitude.
As for the parameters entering the Schechter function (including $M_{UV}^*$) 
we use the observational values in Table 1 of~\citet{Bouwens11}.
The star formation rate (uncorrected for dust attenuation)
for a galaxy with magnitude $M_{UV}$ 
is obtained by manipulation of the expression in~\citet{Madau98} and reads
\begin{equation}
\dot\rho_*(M_{UV})= 5.43\times10^{-(M_{UV}+20)/2.5}~M_\odot \mathrm{yr}^{-1}.
\end{equation}
Eq. (\ref{SFR:eq}) contains contributions up to a limiting
magnitude, $M_{UV}^\mathrm{max}$. \citet{Bouwens11} assumes
$M_{UV}^\mathrm{max}=-10$, corresponding to the mass
scales where galaxy formation is suppressed due to inefficient gas
cooling and/or feedback effects.
Here we generalize this limit to include effects of magnetic pressure
on the Jeans mass. For this purpose we use the filtering mass
$M_F$ \citep{Gnedin98,SchleicherBanerjee08}
\begin{equation}
M_{F}^{2/3}=\frac{3}{a}\int_0^a da' M_g^{2/3}(a')\left[1-\left(\frac{a'}{a}\right)^{1/2}\right],
\end{equation}
where, $M_g$ denotes the maximum of the thermal, $M_J$, and
magnetic Jeans masses, $M_J^B$. These are, respectively
\begin{equation}
M_J=2 M_\odot\left( \frac{c_s}{0.2~\mathrm{km/s}}\right)^3\left( \frac{n}{1000~\mathrm{cm}^{-3}}\right)^{-0.5}
\end{equation}
with $c_S$ the sound speed and $n$ the gas number density, 
and~\citep{Subramanian98p}
\begin{equation}
M_J^B=10^{10} M_\odot \left(\frac{B_0}{3~nG} \right)^3,\label{JeansB}
\end{equation}
with $B_0$ the co-moving field strength.
Finally, to convert the filtering mass scale into a UV magnitude, 
we use the following relation inferred from Fig.~10 of \citet{Salvaterra11}
(which also consistently reproduces the above limiting magnitude in absence of magnetic field)
\begin{equation}
M_{UV}(M_{\mathrm{halo}})=-20-3\times\log_{10}\left (\frac{M_{\mathrm{halo}}}{10^{10.73~M_\odot}}\right).
\end{equation}

\begin{figure}
  \includegraphics[scale=0.49]{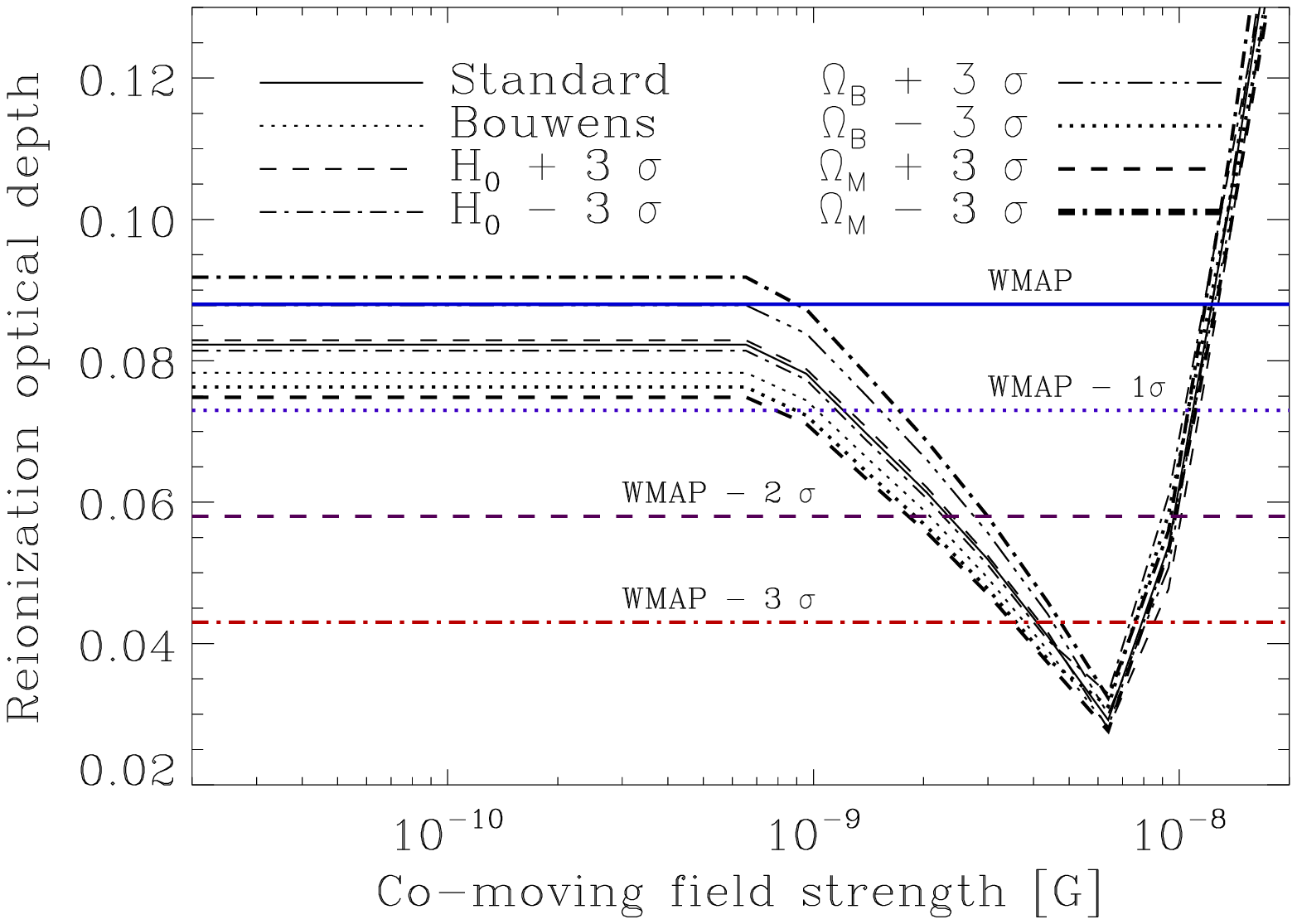}
  \includegraphics[scale=0.49]{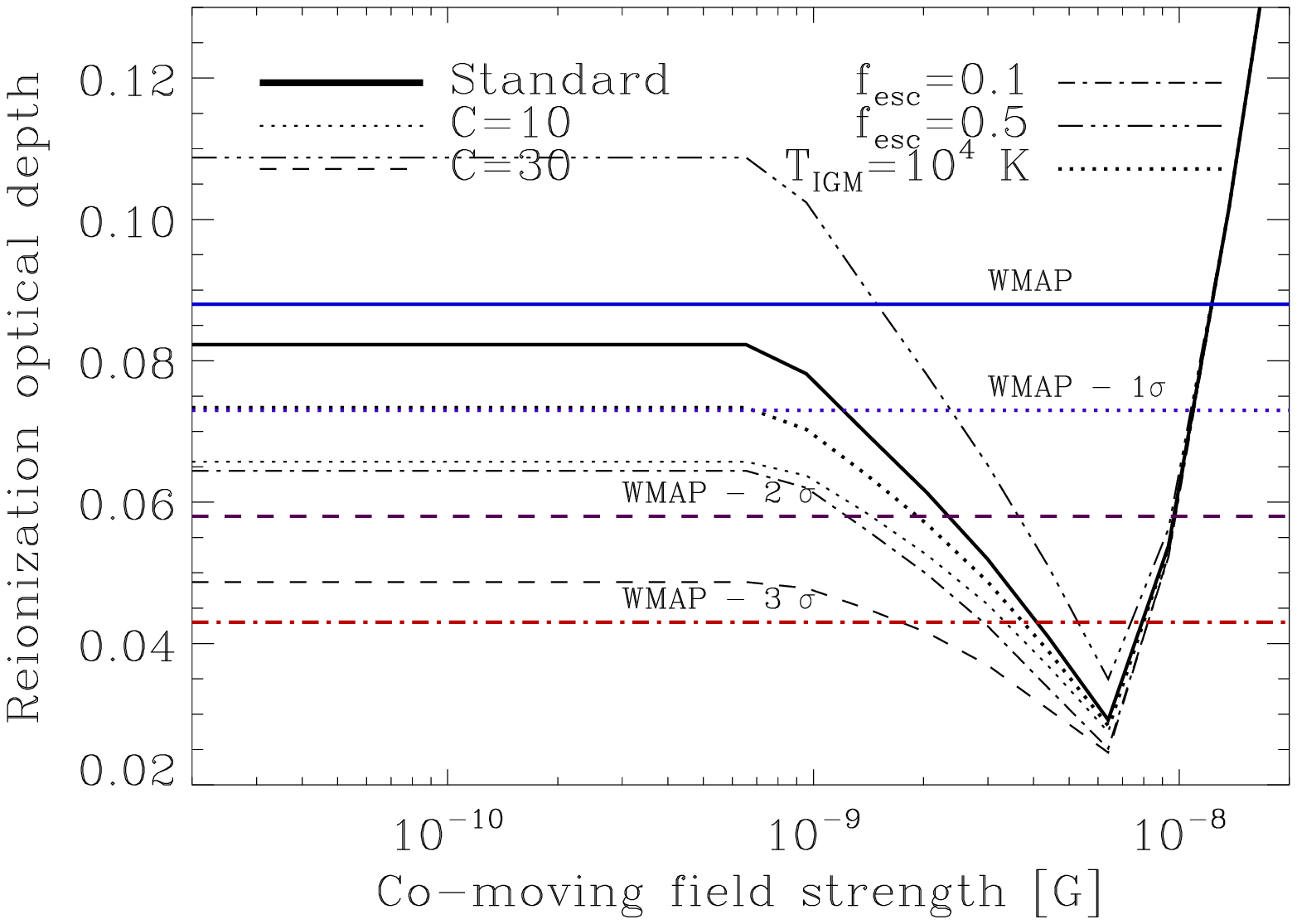}
  \includegraphics[scale=0.49]{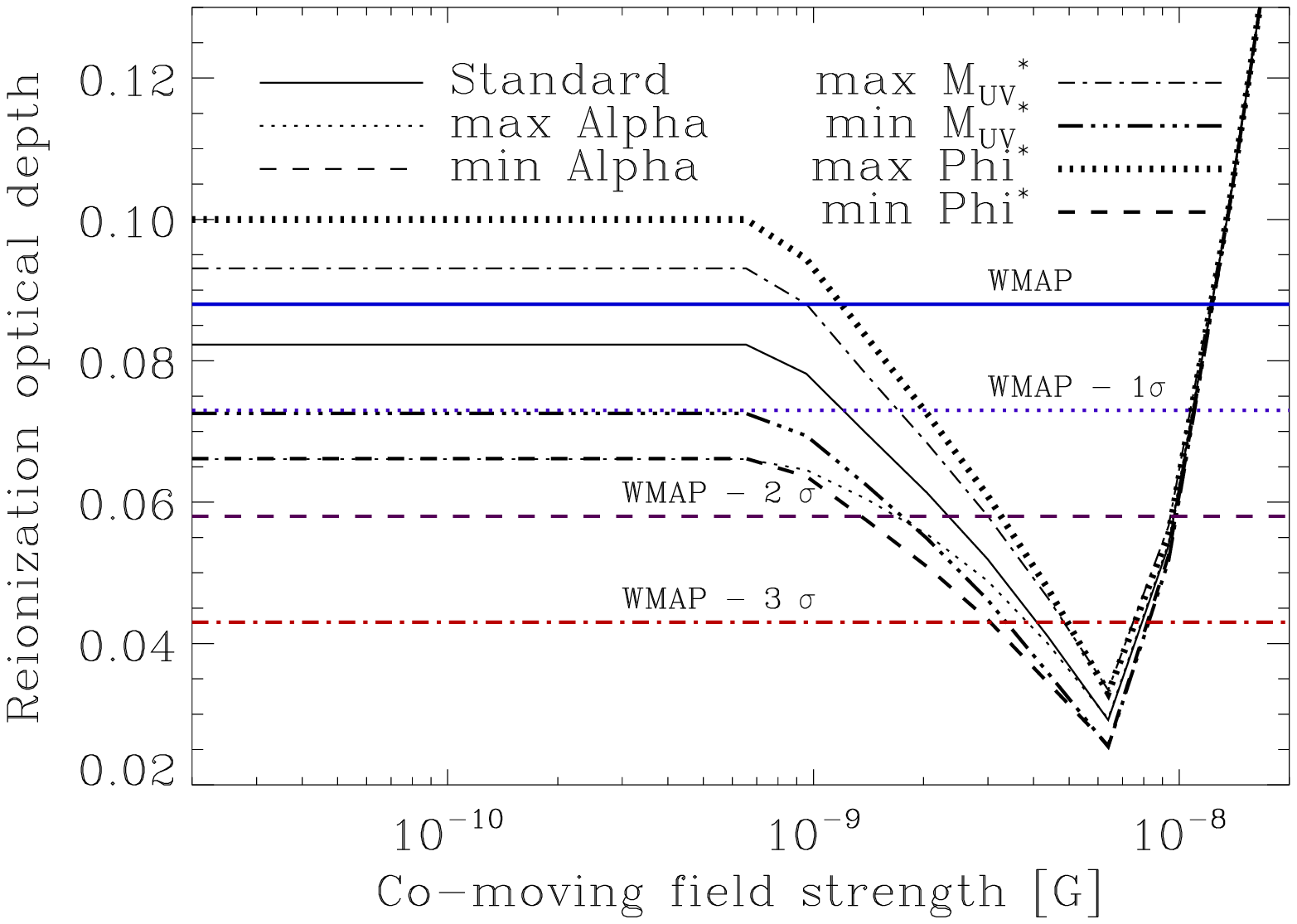}
  \caption{The reionization optical depth as a function of the co-moving field strength. Upper panel: Variation of the cosmological parameters. Mid panel: Variation of the reionization parameters. Lower panel: Variations in the Schechter function. The curves used to derive the upper limit are marked with a thick line.}
  \label{fig:tau}
\end{figure}

\begin{figure}
  \includegraphics[scale=0.49]{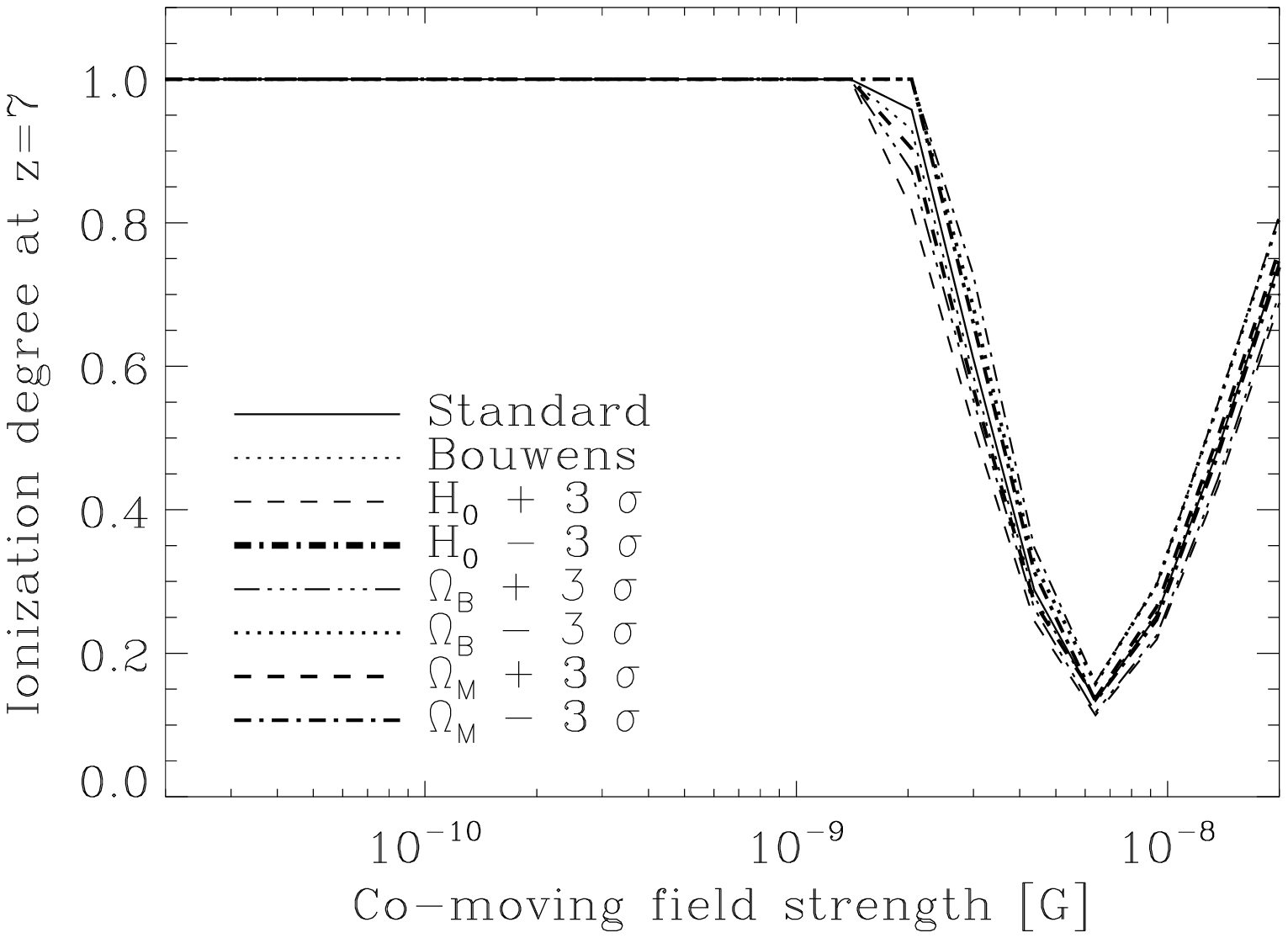}
  \includegraphics[scale=0.49]{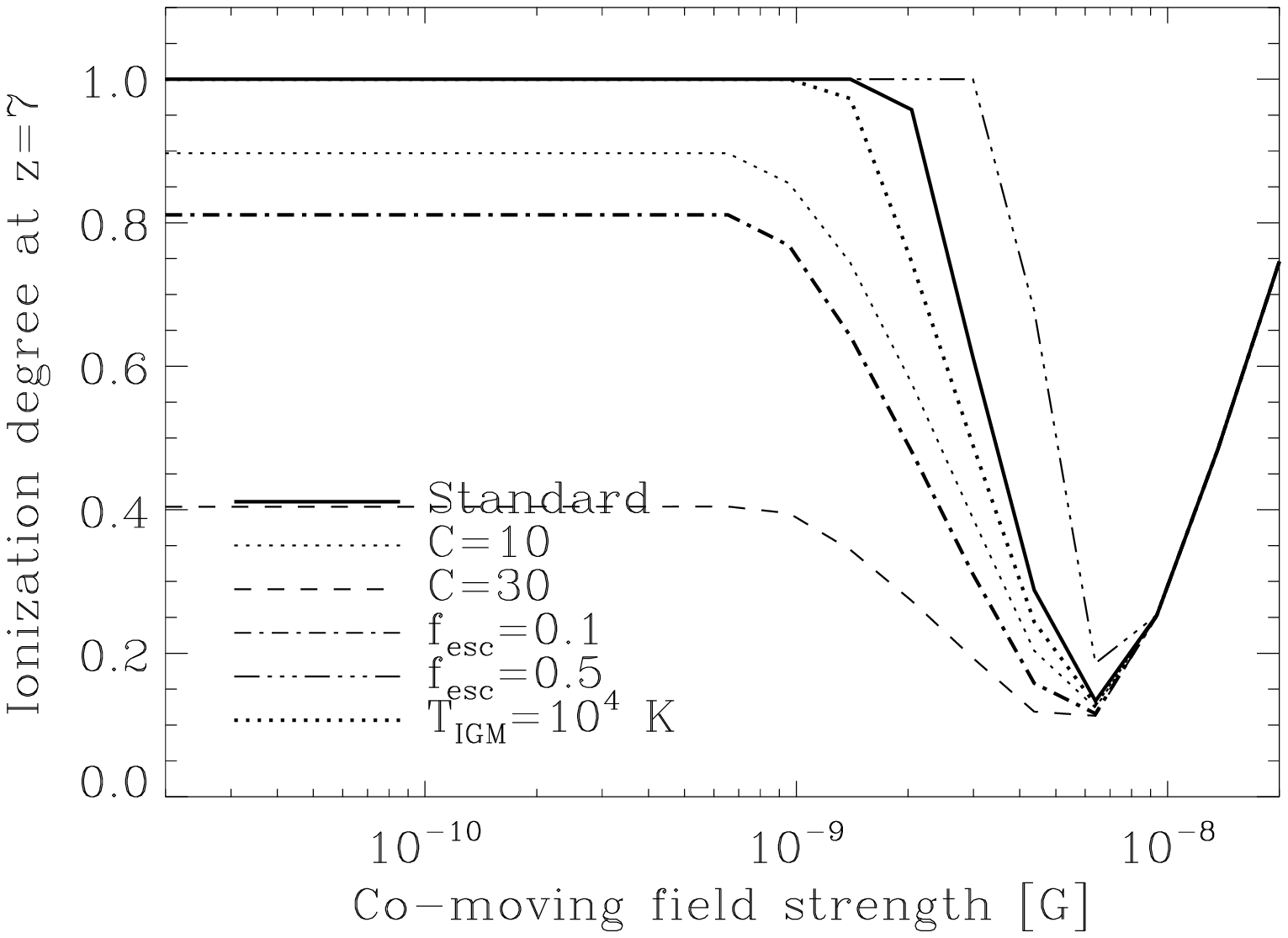}
  \includegraphics[scale=0.49]{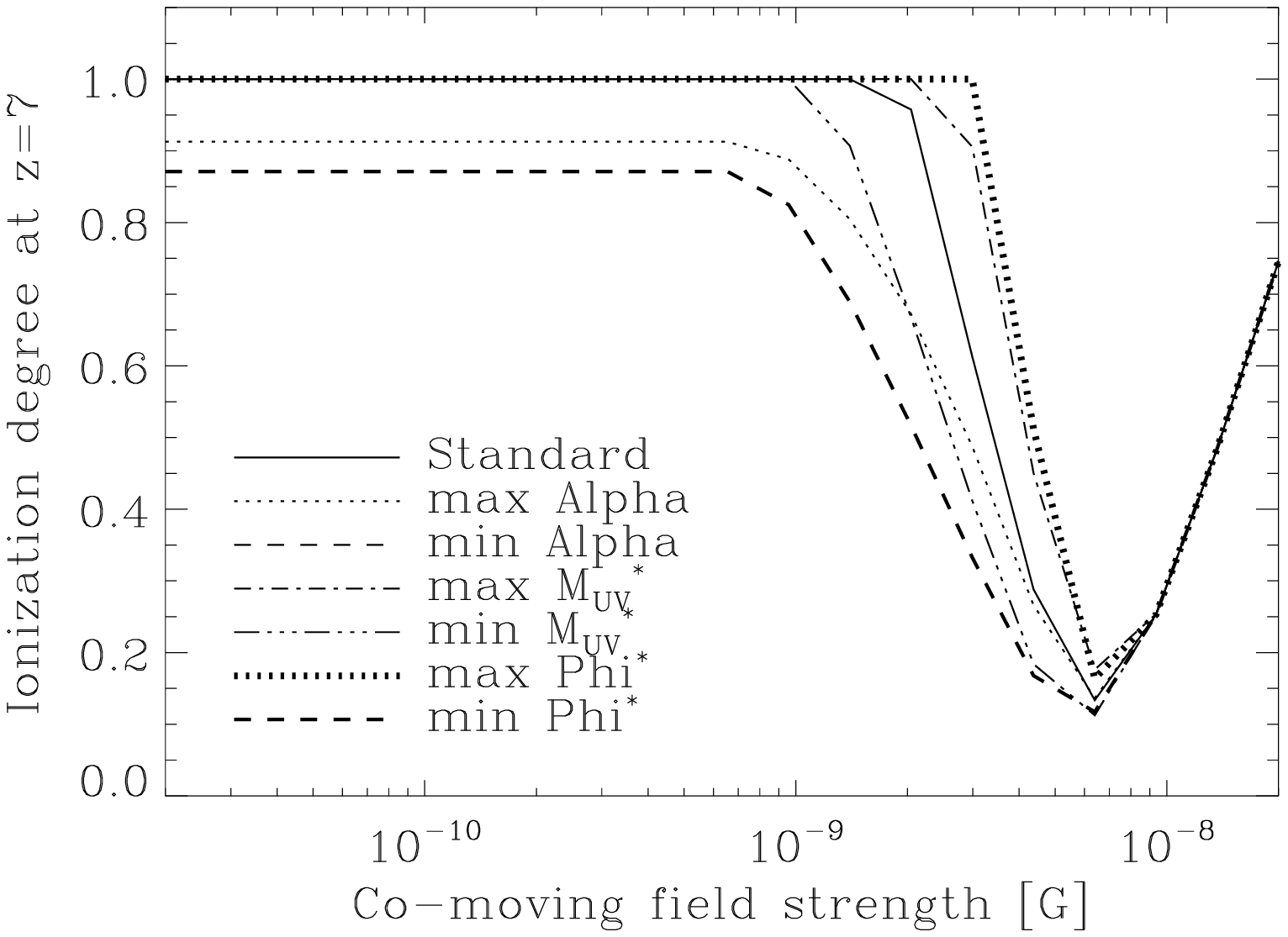} 
 \caption{The ionization degree at $z=7$ as a function of the co-moving field strength. Upper panel: Variation of the cosmological parameters. Mid panel: Variation of the reionization parameters. Lower panel: Variations in the Schechter function. The curves used to derive the upper limit are marked with a thick line.}
    \label{fig:ion}
\end{figure}

In addition to the volume fraction of ionized hydrogen,
our calculation follows the time evolution of
the temperature and ionization degree of the partially-ionized IGM,
and the magnetic field strength.
The full set of equations is given in
\citet{SchleicherBanerjee08} and was implemented in the RECFAST
code\footnote{http://www.astro.ubc.ca/people/scott/recfast.html}
\citep{Seager99}. 
In particular, the magnetic energy decreases with time due to
adiabatic expansion and dissipation into heat through
ambipolar diffusion. For strong fields this can change
substantially the thermal and ionization history of the IGM,
although full reionization cannot be achieved this way, as collisions 
induced by ambipolar
diffusion become inefficient at high temperatures.

With the calculation results, 
the Thompson scattering optical depth is computed as
\begin{equation}
\tau_e=\frac{n_H(0) c}{H_0} \int_{z=0}^{z=z_s}x_{eff}(z)\sigma_T \frac{(1+z)^2}{\sqrt{\Omega_\Lambda+\Omega_m(1+z)^3}}dz,
\end{equation}
where $\sigma_T$ is the Thompson scattering cross section and
 the effective ionization degree is
\begin{equation}
x_{eff}=Q_{HII}+(1-Q_{HII}) x_e
\end{equation}
contributed by both the fully ionized volume fraction as
well as the partially ionized gas.

\begin{figure}
  \includegraphics[scale=0.5]{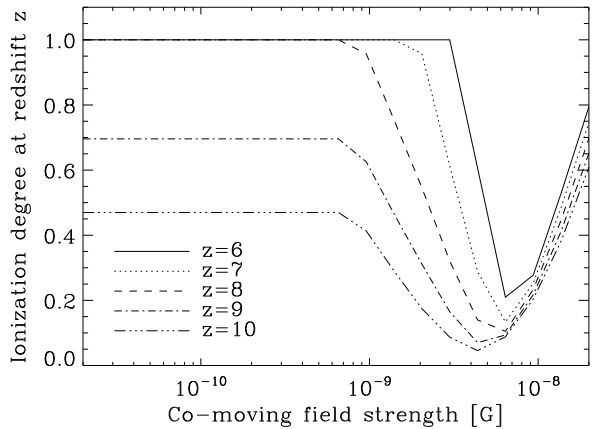}
  \caption{The ionization degree at different redshifts, as a function of the co-moving field strength (in our reference scenario).}
    \label{fig:redshift}
\end{figure}

\section{Results}
In the following, we set the upper limits on the comoving
magnetic field strength, $B_0$, using the Thomson optical depth (a
redshift-integrated quantity) as well as the the observed ionization
degree at $z\sim7$.  We discuss the uncertainties due to the
cosmological parameters, the reionization parameters and the observed
UV luminosity function. Finally, we discuss the possibility of more stringent
constraints from future progress on the high redshift IGM.

\subsection{Constraints from the Thompson optical depth}
Fig.~\ref{fig:tau} shows the constraints on $B_0$ due to the observed
optical depth, $\tau_e$, and its uncertainties. 
Starting from our reference case in the top panel, we see that
for weak magnetic fields the optical depth is 
well within the $1\sigma$ range of WMAP7 results. 
As  $B_0$ increases the low-luminosity cut-off 
of halos contributing ionizing photons is raised and at $B_0\gsim$ nG,
$\tau_e$ starts to drop.  It reaches a minimum at several nG and then,
as collisional ionization resulting from heating due to ambipolar diffusion
kicks in, $\tau_e$ increases 
again~\citep[however, values of several nG for $B_0$, 
would be inconsistent with CMB observations, e.g.,][]{Yamazaki10}.
This is the qualitative trend in all panels of Fig. 1, 2 and 3. 

The top panel also shows the effects of $3\sigma$ variations of the
cosmological parameters, on $\tau_e$. For the purpose of the current
discussion, {\bf higher} values of $H_0$ and smaller values of $\Omega_M$ 
increase the value of $\tau_e$ and thus weaken the constrain on $B_0$.
Thus, the constraint on $B_0$ is set by the highest curve, 
which says that the required field strength at the 95\% CL limit is $3$~nG.
Note also that adopting a cosmology as in \citet{Bouwens11} would have
minor effects on our results.

In the mid panel we consider instead the effect due to 
uncertainties in the reionization parameters.
Such uncertainties can in principle be very large. However, a
large clumping factor $\sim 30$ is inconsistent with the observed 
ionization fraction (see Fig. 2) even in the limit of weak magnetic 
fields. Likewise, a very large escape fraction $f_{esc}\sim 50\%$ produce
too high an ionization rate at 
$z\lsim 6$~\citep{Srbinovsky2010PASA...27..110S,mfwb04}.
With these considerations, a meaningful
95\% CL upper limit of 2nG is implied by our results.

Finally, in the bottom panel we consider the uncertainties due to
3$\sigma$ observational errors in the parameters $\alpha$, 
$M_{\mathrm{UV}}^*$ and $\Phi^*$, of the Schechter function.
{Here, it is large values of $\Phi^*$ or $M_{UV}^*$ that
increase $\tau_e$, and therefore  weaken our constraint.}
Thus, the highest curve on the plot shows that the computed value of $\tau_e$ 
remains within $2\sigma$ of the WMAP7 result for $B_0\lsim$3nG.
Note that the redshift evolution of the Schechter function 
parameters is also very uncertain, but we have verified that it does
not lead to more significant uncertainties on the current constraints.
We thus arrive at a 95\% CL upper limit of $\sim 2-3$~nG on $B_0$.

\subsection{Constraints from the ionization degree}
Fig.~\ref{fig:ion} shows the constraints due to the observed
ionization degree, $x_{\mathrm{eff}}$, and its uncertainties.
Following the analysis
of~\citet{Bolton11}, the IGM is assumed mostly ionized at $z\sim7$.
Note that for the present purposes the accurate value of the
ionization degree is not crucial and even a 90\% value would lead to
very strong constraints independent of the specifically adopted CL.
In our reference case (top panel), 
we find at $z=7$ an ionization degree close to 1
for co-moving field strengths up to $\sim1.5$~nG. For increasing field
strengths, $x_{\mathrm{eff}}$ drops rapidly, as the magnetic Jeans mass
increases with $B_0^3$, significantly suppressing the production of
ionizing photons. Thus considering all the uncertainties due to the 
cosmological parameters leads to $B_0<2$nG.

As already discussed for the optical depth, the
uncertainties due to the reionization parameters are considerably
higher (mid panel). However, cases with high clumping factors, low escape
fractions or low IGM temperatures and even very high escape 
fraction~\citep{Srbinovsky2010PASA...27..110S,mfwb04}
are inconsistent even in the limit
of weak magnetic fields. For the remaining cases, 
the ionization degree drops rapidly for $B_0$ above a nG,
and the most conservative upper limit 
corresponds again to $B_0\lsim $ 2nG.

As for the uncertainties in the parameters of the Schechter function
bottom panel of Fig.~\ref{fig:ion} shows that some extreme values 
can be ruled out because inconsistent with the ionization degree at
$z=7$, even with negligible $B_0$.
However, the most important source of uncertainty for the upper limit 
on $B_0$ are those related to $M_{UV}^*$. This forces our constraint
to $B_0\lsim$ 3nG.
This constraint will be improved once the Schechter function is observed more
accurately. Again, we checked that the uncertainties in the redshift
evolution have no impact on our results.From the observed ionization degree, we thus find an overall
constraint $B_0\lsim 2-3$~nG.

In Fig.~\ref{fig:redshift} we show the ionization degree at
different redshifts as a function of the co-moving field strength.
The plot illustrates the sensitivity of our constraint to the redshift
at which the ionization degree is measured. In particular our 
constraint would be slightly weaker if based on data at $z=6$, while it
would considerably improve if a high ionization degree at
$z=8$ could be observationally established. At even higher redshifts,
our model predicts the presence of a substantial neutral fraction in
the IGM which could be translated into a further independent constraint.

\section{Discussion and outlook}
Using the combined constraints from the observed Thomson optical
depth, the ionization degree of the IGM at $z=7$ and the observed UV
luminosity function at high redshifts, we have derived robust upper
limits of $2-3$~nG, virtually independent of the CL, on the strength
of primordial magnetic fields.  Previous work~\citep[e.g.][see also
Introduction]{Sethi05,Tashiro06a,SchleicherBanerjee08} found similar
limits but generally was based on the Thomson optical depth only, used
less refined physical models, had a lower CL. However, the similarity of
conclusion also suggests that we are approaching the intrinsic limitation
of the methods.

Our results are mostly based on the lower cutoff of the luminosity
function of galaxies contributing ionizing photons, set by the
magnetic Jeans mass (Eq.~\ref{JeansB}).
On the other hand, heating effects due to ambipolar diffusion
are important for $B_0$ close to 10 nG, which is ruled out by CMB observations.
Our constraints on the primordial magnetic field would be mostly 
strengthened
by measurements of a high ionization degree at redshift $z\gsim 8$.
This is simply because the earlier the IGM is reionized, the more important
becomes the contribution of ionizing photons from halos whose collapse
is hindered by primordial magnetic fields.
However, improvements beyond an order of magnitude are likely out of
reach, as the formation of structure at the magnetic Jeans mass
corresponding to 10$^{-10}$G (see Eq.~\ref{JeansB}) is severely
affected by other feedback processes~\citep[see, e.g., discussion
in][]{Bouwens11}.
Other limiting factors in the derivation of our constraints are mostly
due to uncertainties in the UV luminosity function at high redshift
and lack of information about the IGM ionization fraction at redshift
higher than 7. There is reason to expect good progress in
these areas in the near future. For example, JWST\footnote{Homepage
  JWST: http://www.stsci.edu/jwst/} will provide a unique opportunity
to probe the IGM at and beyond redshift $10$, Lyman $\alpha$ emitters
may provide additional information on the strength of the
photoionizing background \citep[e.g.][]{Latif11b}, while the Planck
satellite\footnote{Homepage
  Planck:http://www.rssd.esa.int/index.php?project=Planck} will
provide an improved measurement of the Thomson optical depth. These additional data may help to improve our understanding both with respect to reionization and the origin of magnetic fields.

\section*{Acknowledgments}
We thank an anonymous referee for constructive comments and
R. Banerjee, B. Ciardi, C. Federrath, D. Galli, R.S. Klessen and
R. Salvaterra for stimulating discussions.

\bsp


\label{lastpage}

\end{document}